\documentstyle[aps,multicol,epsf,epsfig]{revtex}

\begin{document}

\draft

\title{Monochromatic path crossing exponents and graph connectivity
       in 2D percolation}

\author{Jesper Lykke Jacobsen and Paul Zinn-Justin\thanks{Electronic addresses: \{jacobsen,pzinn\}@lptms.u-psud.fr}}

\address{Laboratoire de Physique Th\'eorique et Mod\`eles Statistiques \\
         Universit\'e Paris-Sud, B\^atiment 100, 91405 Orsay Cedex, France}

\date{June 2002}

\maketitle

\begin{abstract}

We consider the fractal dimensions $d_k$ of the $k$-connected part of
percolation clusters in two dimensions, generalizing the cluster ($k=1$)
and backbone ($k=2$) dimensions. The codimensions $\tilde{x}_k=2-d_k$
describe the asymptotic decay of the probabilities
$P(r,R) \sim (r/R)^{\tilde{x}_k}$ that an annulus of radii $r \ll 1$
and $R \gg 1$ is traversed by $k$ disjoint paths, all living on the
percolation clusters. Using a transfer matrix approach,
we obtain numerical results for $\tilde{x}_k$,
$k\le 6$. They are well fitted by the Ansatz
$\tilde{x}_k = \frac{1}{12}k^2 + \frac{1}{48}k + (1-k)C$, with
$C = 0.0181 \pm 0.0006$.

\end{abstract}

\begin{multicols}{2}

Percolation is a classical model of statistical mechanics
\cite{Stauffer94,Bunde96}, and plays an important role in the study of
disordered systems \cite{Gruzberg}. It is also one of the simplest models
displaying a critical point. In two dimensions, exact values for a variety of
critical exponents have been found over the last two decades, and quite
recently many of them have been confirmed by rigorous probabilistic arguments
\cite{Smirnov01}. Most of these exponents can be defined through the fractal
dimensions of suitably defined sets at the percolation threshold.

In the present Letter we shall be concerned with an infinite family of
critical exponents whose exact values remain unknown to this date.
These exponents characterize the connectivity structure of the percolating
cluster(s) at criticality.

Following Tutte \cite{Tutte84}, we define a graph to be $k$-connected (for $k
\ge 1$) if no separation into disconnected subgraphs is possible by
eliminating at most $k-1$ vertices along with their ingoing edges. It is easy
to see that we may equivalently require any two vertices in the graph to be
connected by (at least) $k$ disjoint paths. 1-connected
graphs are simply the percolation clusters, and to inquire into
the connectivity structure of a given cluster, we may decompose it into its
largest 2-connected components \cite{Tutte84}
(better known as 2-blocks, or ``blobs'', in the
percolation literature), and so on. Tutte has shown that the decomposition of
a 2-connected graph into its largest 3-blocks is unique \cite{Tutte84}.
3-blocks are relevant for applying Kirchhoff's laws to resistor networks,
and are useful for analyzing the performance of certain algorithms
\cite{Paul02}.

To better study the transport properties of percolation clusters, we
henceforth consider critical percolation in a large square of linear
size $L$, and we specialize to clusters that connect to the boundary
of the system. In the limit $L\to\infty$, the boundary becomes the
``point at infinity'', and the above definition states that a given vertex
is $k$-connected if it is connected to infinity by (at least) $k$
disjoint paths within a percolating cluster. Following Ref.~\cite{Paul02},
we shall call the set of $k$-connected vertices the $k$-bone.
The 2-bone is of course nothing but the (geometrical) backbone, i.e.~the
part of the percolation cluster that sustains a non-zero current, when
a voltage difference is applied between its two terminal points.
(As usual we disregard rare Wheatstone's bridge-like arrangements.)

We shall here be interested in the fractal dimension $d_k$ of the $k$-bone,
assuming its mass to change with system size like $L^{d_k}$. The cluster
dimension $d_1=\frac{91}{48}$ has been known exactly for a long time
\cite{denNijs79,Nienhuis80,Lawler01}. The backbone dimension has recently been
related to the solution of a partial differential equation \cite{Lawler01},
which however appears to be intractable, even numerically. Still, numerical
estimates are available from Monte Carlo \cite{Grassberger99} and transfer
matrix methods \cite{JZJ}: $d_2 = 1.6431 \pm 0.0006$. After the completion of
this work, a first estimate for $d_3$ appeared: $d_3 = 1.2 \pm 0.1$
\cite{Paul02}. Actually, this result was obtained from block-decomposition of
clusters and backbones, but for reasons of universality we expect it to apply
to the 3-bone as well. Also, the above definitions are stated for site
percolation, but the exponents should be the same for bond percolation, with
the clusters being separated by cutting edges rather than vertices.

\begin{figure}
\centerline{
\epsfxsize=4.0cm
\epsfclipon
\epsfbox{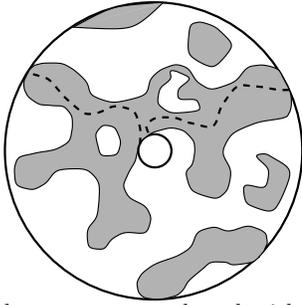}
}
\caption{Annular geometry endowed with critical percolation (here shown in
the continuum limit). The existence of two disjoint traversing paths on
the clusters implies that this configuration contributes to the 2-bone.}
\label{fig:annulus}
\end{figure}

A useful alternative formulation of the $k$-bone problem is obtained by
passing to an annular geometry, limited by two concentric circles of
radii $r \ll 1$ and $R \gg 1$, by means of a conformal mapping.
(This is permissible since percolation has been proved
to be conformally invariant \cite{Smirnov01}.) Interpreting the inner
circle as the point which is a potential element of the $k$-bone, and
the outer as the point at infinity, we see that a given percolating
configuration in the annulus contributes to the $k$-bone if and only if
the two circles are connected by $k$ disjoint paths on the percolating
cluster(s); see Fig.~\ref{fig:annulus}. The fractal dimension $d_k$ of the
$k$-bone is linked to the scaling of the path-crossing probability
$P_k(r,R) \sim (r/R)^{\tilde{x}_k}$ through the
codimension $\tilde{x}_k = 2-d_k$ \cite{Aizenman}.

A more general class of path-crossing exponents can be defined from
traversing configurations where some of the paths live on the percolating
clusters (black paths), and the rest on the {\em dual} clusters (white paths).
Interestingly, the corresponding critical exponents $x_k$ only depends on
the number of paths, $k \ge 2$, and not on their colors, provided that
both are represented \cite{Aizenman}. Their values
\begin{equation}
 x_k = \frac{1}{12}(k^2-1)
 \label{polychrom}
\end{equation}
are known rigorously \cite{Aizenman} and differ from those of the monochomatic
exponents $\tilde{x}_k$ defined above.

Some information on the $\tilde{x}_k$ is provided by the inequalities
\begin{equation}
\tilde{x}_k < x_{k+1}.
 \label{inequality}
\end{equation}
This inequality is valid since a configuration contributing to
$x_{k+1}$ can be taken to have $k$ black paths and one white path. Clearly,
it then also contributes to $\tilde{x}_k$. Furthermore, as $k\to\infty$,
the effect of a single extra path should be small and we expect
it to be of the same order if one changes the color of one of the
$k$ existing paths; thus, $x_{k+1}-\tilde{x}_k\approx
x_{k+1}-x_k=O(k)$.

In view of Eq.~(\ref{polychrom}) we find the
asymptotic result $\tilde{x}_k = \frac{1}{12}k^2+O(k)$ as $k\to\infty$.
In analogy with Eq.~(\ref{polychrom}) it thus seems natural to conjecture
that the spectrum $\tilde{x}_k$ is quadratic in $k$. From
$\tilde{x}_1 = \frac{5}{48}$ we then get
\begin{equation}
 \tilde{x}_k = \frac{1}{12} k^2 + \frac{1}{48}k + (1-k)C,
 \label{spectrum}
\end{equation}
with $C = 0.0181 \pm 0.0006$ from the numerical result on $\tilde{x}_2$
\cite{JZJ}.

To check the conjectured form of the spectrum, Eq.~(\ref{spectrum}), we
now turn to our numerical results. But first we must briefly describe
our transfer matrix algorithm; it is a natural
generalization of the one used in
\cite{JZJ}.

First we consider the annulus of Fig.~\ref{fig:annulus} as
a cylinder with circular space and radial time.
We have tried several choices of lattices to discretize it.
For practical applications, it turns out to be 
best (see \cite{JZJ}) to use a square lattice with a ``light-cone'' 
orientation, that is such that
the periodic direction forms a 45 degrees' angle with the two
axes
of the lattice. We then define the discrete time slices
such that they intersect the lattice at vertices only: we call
$L$ the number of such vertices (in units of the lattice spacing the period
is then $L\sqrt{2}$).

\begin{figure}
\centerline{
\epsfxsize=6.0cm
\epsfclipon
\epsfbox{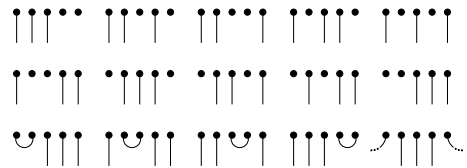}
}
\caption{Path configurations for $k=3$, $L=5$.}
\label{fig:paths}
\end{figure}

Next we define the basis of states on which our transfer matrix acts.
A basis state is a {\it collection}\/ of path configurations; a path
configuration is the data of the positions of our $k$ paths at a given
time, with possible additional ``arches'' to allow backtracking of
paths, see Fig.~\ref{fig:paths}. Note that the encoding of states can
be easily implemented as follows: basis states are encoded as
sorted lists of path configuations, and path configurations are
represented by words of length $L$ made out of the four letters
$\{ opening,closing,path,empty \}$ and which
contain $k$ letters $path$. The letters $opening$ and $closing$ define
the backtracking arches.

The transfer matrix itself acts on a basis states
by ``evolving'' all the path configurations it contains with
a single configuration of bonds (i.e.\ percolating/non-percolating
state of each bond) and recombining
the result into a single state, then summing over all configurations
of bonds. Evolving a path configuration means considering all possible
continuations of the paths from one time slice to the next,
including possible appearances of new arches, or existing arches connecting
to paths or other arches. We use sparse matrix factorization techniques to
build up the entire transfer matrix; dihedral symmetries are quotiented over
once a time slice has been completed.

As in \cite{JZJ}, it is convenient to allow ``perpendicular tangencies'' for
paths even though they should be in principle excluded; that is to allow two
paths to touch at one vertex in the configuration whenever the tangent of
either path is perpendicular to the transfer direction (we still exclude
``parallel tangencies''). It is expected, and we have numerically verified,
that inclusion of either or both types of tangencies and/or changing the
lattice orientation does not alter the first finite-size correction of the
eigenvalue of the transfer matrix, which yields $\tilde{x}_k$. However, the
particular choice of allowing perpendicular tangencies on the light-cone
oriented lattice has the advantage of greatly decreasing the number of states
generated (see below) and improving the convergence properties of the
finite-size data.

Finally, the following procedure is used to compute the matrix elements of the
transfer matrix: Starting with an arbitrary basis state (e.g.\ the one
consisting of the single path configuration $path^k empty^{L-k}$),
one acts on it with
the transfer matrix, stores the corresponding matrix elements, then considers
all new basis states generated and iterates the procedure until no further new
states are generated \cite{footnote}. What we build this way is a submatrix of
the transfer matrix corresponding to a stable subspace; this submatrix is in
fact much smaller than the full transfer matrix, which is essential for
practical applications. One then extracts its largest eigenvalue
$\lambda_k(L)$, which yields the codimension $\tilde{x}_k$ via the formula
\begin{equation}
{1\over 2^{2L}} \lambda_k(L)=1-{\pi \tilde{x}_k\over L}+o(L^{-1})
\end{equation}

We show in Tab.~\ref{tab:eig} the results for $2\le k\le 6$. Since the memory
and time requirements presumably grow factorially with $L$, we cannot push the
calculation very far in $L$. It is however sufficient to estimate
$\tilde{x}_k$, which is also given in Tab.~\ref{tab:eig} together with
approximate error bars. The data for $k=2$ are in fact taken from \cite{JZJ},
where the state space was reduced by exploiting that the case of two paths can
be treated as an extra backtracking arch.

For $k>2$, all the exponents $\tilde{x}_k$ are consistent with, but less
precise than, the conjectured spectrum (\ref{spectrum}) with $C$ given by
$\tilde{x}_2$. We also note that the inequality $x_k < \tilde{x}_k$ seems to
be satisfied, and if would be interesting if one could prove this. Another
open question is the possible rationality of the $\tilde{x}_k$; in this
respect, the failure of computing these exponents by conformal field theory is
particularly intriguing.

For $k \ge 5$, the dimensions $d_k=2-\tilde{x}_k$ are negative.
Physically this means that $k$-bones with $k \ge 5$ become increasingly
rare as the system size increases. Of course, the lattice model does not
support a $k$-block when $k$ exceeds the coordination number of the lattice.
However, the exponents $\tilde{x}_k$ characterize the continuum limit,
and are thus believed to be independent of the microscopic details;
an appropriate lattice definition of the $k$-bone for high $k$ is obtained
by demanding that $k$ independent paths connect any small neighborhood to the
point at infinity.

Finally, we remark that the $k$-bone problem extends to the Kasteleyn-Fortuin
representation \cite{KF69} of the $q$-state Potts model (bond percolation
being the limit $q\to 1$). Our numerical algorithm can straightforwardly
be adapted to this case as well.

\subsubsection*{Acknowledgements}

We thank S.~Kirkpatrick and J.~Vannimenus for useful discussions.

\end{multicols}

\begin{table}%[eig]
\caption{Eigenvalues $\lambda_k(L)$ of the transfer matrix 
and estimate of $\tilde{x}_k$ for $2\le k\le 6$.}
\begin{tabular}{cccccc}
$k$&2&3&4&5&6\\ \hline
$L$\\
4& 0.718747415570&0.413598206498&0.121093750000\\
5& 0.775012703547&0.526618869796&0.257122218539&0.061523437500\\
6& 0.812529692986&0.603476157424&0.362299981029&0.153371684616&0.031005859375\\
7& 0.839330907375&0.658986646726&0.443031423565&0.237989873966&0.088905009155\\
8& 0.859432882632&0.700919030179&0.506272495802&0.310651059489&0.150977532764\\
9& 0.875067710677&              &0.556925756584&0.372225212770&0.210050339522\\
10&&              &              &              &0.263993624780\\
\hline
$\tilde{x}_k$&$0.3569\pm0.0006$&$0.77\pm0.02$& $1.33\pm0.03$& $2.1\pm0.2$& $3.0\pm0.3$\\
\end{tabular}
\label{tab:eig}
\end{table}

%%%%%%%%%%%%%%%%%%%%%%%%%%%%%%%%%%%%%%%%%%%%%%%%%%%%%%%%%%%%%%%%%%%%%

\begin{references}
\bibitem{footnote} This stage of the computations was parallelized on a
 network of computers (one server and a dozen of clients) linked by
 socket-type communication protocols.
\end{references}
\end{document}